\documentclass{jetpl}
\input epsf

\twocolumn
\lat
\title{BL Lacertae are probable sources of the observed ultra-high 
energy cosmic rays}
\rtitle{BL Lacertae are probable sources of \ldots}
\sodtitle{BL Lacertae are probable sources of the observed ultra-high 
energy cosmic rays}

\author{P.G.~Tinyakov$^{a,c}$ and I.I.~Tkachev$^{b,c}$ }
\rauthor{P.\,G.\,Tinyakov and I.\,I.\,Tkachev}
\sodauthor{Tinyakov, Tkachev}

\address{$^a${\it Institute of Theoretical Physics,\\
University of Lausanne, CH-1015 Lausanne, Switzerland\\}
$^b${\it CERN Theory Division, CH-1211 Geneva 23, Switzerland\\}
$^c${\it Institute for Nuclear Research of the 
Academy of Sciences of Russia, Moscow 117312, Russia }}

\abstract{ We calculate angular correlation function between
ultra-high energy cosmic rays (UHECR) observed by Yakutsk and AGASA
experiments, and most powerful BL Lacertae objects. We find
significant correlations which correspond to the probability of
statistical fluctuation less than $10^{-4}$, including penatly for
selecting the subset of brightest BL Lacs. We conclude that some of BL
Lacs are sources of the observed UHECR and present a list of most
probable candidates.  }

\PACS{98.70.Sa}

\begin{document}

\maketitle

\paragraph*{Introduction.} 

Identification of sources of ultra-high energy cosmic rays (UHECR) is
extremely important. Knowing production sites of UHECR will help to
explain the apparent absence of the Greisen-Zatsepin-Kuzmin (GZK)
cutoff \cite{GZK} by selecting a particular class of models. In the
case of astrophysical origin it will give an invaluable information on
physical conditions and mechanisms which may lead to acceleration of
particles to energies of order $10^{20}$ eV. In the case of
extragalactic origin, it will provide a direct information about
poorly known parameters which influence propagation of UHECR, such as
extragalactic magnetic fields and universal radio background.

There are observational reasons to believe that UHECR are produced by
compact sources. It has been known for quite a while that the observed
highest energy cosmic rays contain doublets and triplets of events
coming from close directions
\cite{clusters1,Takeda:1999sg,Uchihori:2000gu}. 
Our recent analysis \cite{Tinyakov:2001ic} based on the calculation of
angular correlation function shows that explanation of clusters by
chance coincidence is highly improbable: the correlation function for
Yakutsk events \cite{yk} with energies $E>2.4\times 10^{19}$~eV has an
excess at $4^{\circ}$ which would occur with probability $2\times
10^{-3}$ for the uniform distribution, while the correlation function
for AGASA events \cite{Takeda:1999sg,Hayashida:1999zr} with energies
$E>4.8\times 10^{19}$~eV has an excess at $2.5^{\circ}$ corresponding
to chance probability $3\times 10^{-4}$. The combined probability of
the fluctuation in both sets is $4\times10^{-6}$. So significant
autocorrelations should imply also large correlation of these events
with their actual sources. It is a purpose of the present paper to
identify these sources.

The clustering of UHECR by itself imposes certain constraints on
possible source candidates. With the observed fraction of events in
clusters, the total number of sources can be estimated along the lines
of Ref.~\cite{Dubovsky:2000gv} to be of order several hundred.  If the
GZK cutoff is absent (or at energies below the cutoff), this estimate
gives the number of sources in the entire Universe. Thus, to produce
observed clustering, the extragalactic sources have to be extremely
rare as compared to ordinary galaxies.  Taking $10^3$ uniformly
distributed sources for an estimate, the closest one is at $z\sim
0.1$.

Various astrophysical candidates such as neutron stars, supernovae,
gam\-ma-ray bursts, colliding galaxies, active galactic nuclei (AGN),
lobes of radio-galaxies, dead quasars and others (for a review see
Refs.\cite{BSigl} and references therein) have been proposed as
sources of UHECR. Possible connection of highest-energy cosmic rays
with these objects was considered in
Refs.~\cite{Takeda:1999sg,Uchihori:2000gu,correlations}. In this paper
we study correlations of UHECR with BL Lacertae (BL Lac) objects which
comprise a subclass of AGN. Our motivations for selecting BL Lacs
are as follows. If AGNs are sources, only those which have jets
directed along the line of sight, or blazars, can correlate with
observed UHECR events (regardless of the distance to a blazar in a
world without GZK cut-off), since particles accelerated in a
relativistic jet are strongly beamed. Blazars include BL Lacs and
violently variable quasars with flat and highly polarized
spectra. These spectral features give direct indication of seeing a
relativistically beamed jet very close to the line of sight.  BL
Lacs is a subclass of blazars characterized, in addition to the
above spectral features which they share, by the (near) absence of
emission lines in the spectra. This very important distinction
indicates low density of ambient matter and radiation and, therefore,
more favorable conditions for acceleration to highest energies.

The most recent catalog of AGNs and quasars contains 306 confirmed BL
Lacs \cite{catalogue}.  While this is the richest catalog we are aware of,
it still may be incomplete. However, this
is not crucial for establishing correlations between BL Lacs and UHECR
events.  Correlations of BL Lacs with UHECR were not studied
before. We show that these correlations do exist and are statistically
significant.

\paragraph*{Method and results.}
Our method is based on calculation of the angular correlation function
and is similar to the one we have used in Ref.~\cite{Tinyakov:2001ic}.
For each BL Lac, we divide the sphere into concentric rings (bins)
with fixed angular size. We count the number of events falling into
each bin and then sum over all BL Lacs, thus obtaining the numbers
$N_i$ (data counts). We repeat the same procedure for a large number
(typically $10^6$) of randomly generated sets of UHECR events.  This
gives the mean Monte-Carlo counts $N^{\rm MC}_i$, the variance
$\sigma^{\rm MC}_i$ and the probability $p(\delta)$ to match or exceed
the data count observed in the first bin. This probability is a
function of the bin size $\delta$. Peaks of $(N_i-N_i^{\rm
MC})/\sigma_i$ or minima of $p(\delta)$ with respect to $\delta$ show
angular scales at which correlations are most significant.

The Monte-Carlo events are generated in the horizon reference frame
with the geometrical acceptance
\[
dn \propto \cos
\theta_z \sin \theta_z d\theta_z, 
\]
where $\theta_z$ is the zenith angle. Coordinates of the events are
then transformed into the equatorial frame assuming random arrival
time. This transformation depends on the latitude of the experiment,
so events simulating different experiments are generated
separately. The distribution of the generated Monte-Carlo events 
in declination and right ascension reproduces well that of 
the experimental data.

We have shown in Ref.~\cite{Tinyakov:2001ic} that autocorrelations are
most significant for the two sets of UHECR events: 26 Yakutsk events
with energy $E>2.4\times 10^{19}$~eV and 39 AGASA events with energy
$E>4.8\times 10^{19}$~eV. If BL Lacs are sources of UHECR, their
correlations with UHECR should be particularly large for these two
sets.  Assuming that energies of the events are not important for
correlations at small angles, we combine them together in one set of
65 events.

\begin{figure}
\begin{center}
\leavevmode\epsfxsize=3.4in\epsfbox{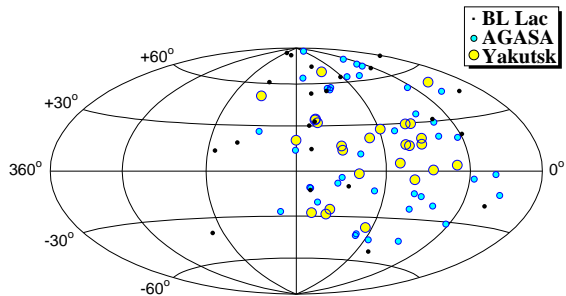}
\end{center}
\caption{FIG. 1. The sky map (in Galactic coordinates) with 65 UHECR events 
(circles) and BL Lacertae objects with cuts (\ref{cuts1}).}
\label{fig:skymap}
\end{figure}

Since acceleration of particles to energies of order $10^{20}$~eV
typically requires extreme values of parameters, probably not all BL
Lacs emit UHECR of required energy. We assume that this ability is
correlated with optical and radio emissions, and select the most
powerful BL Lacs by imposing cuts on redshift, apparent magnitude and
6~cm radio flux.  
For more
than a half of BL Lacs the redshift is not known. It is generally
expected that these BL Lacs are at $z>0.2$. We include them in the
set. The cuts
\begin{equation}
z>0.1 \mbox{~or unknown;~~} {\rm mag} <18 ;~~
F_6>0.17\mbox{~Jy} 
\label{cuts1}
\end{equation}
leave 22 BL Lacs which are shown in Fig.~\ref{fig:skymap} together
with 65 cosmic rays from the combined set. The dependence on cuts is
discussed below.

\begin{figure}
\begin{center}
\leavevmode\epsfxsize=3.4in\epsfbox{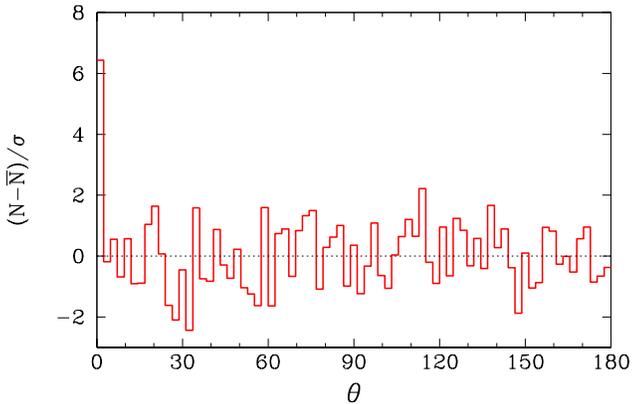}
\end{center}
\caption{FIG. 2. The angular correlation function between the combined set of 
UHECR and BL Lac set (\ref{cuts1}).}
\label{fig:corr2.5}
\end{figure}

As one can see from Fig.~\ref{fig:skymap}, two of 22 BL Lacs coincide
with the two triplets of UHECR events, one coincides with a doublet
and two BL Lacs lie close to single events. This is reflected in the
correlation function, which is plotted in Fig.~\ref{fig:corr2.5} for
the bin size $2.5^{\circ}$. It has 8 events in the first bin while
1.25 is expected for the uniform distribution. The probability of such
an excess is $2\times 10^{-5}$. BL Lacs and UHECR events which
contribute to this correlation are listed in Table~1. Note that at
large angles the correlation function fluctuates around zero, which
shows that the acceptance in the Monte-Carlo simulation is chosen
correctly.

\begin{table}
\caption{Table 1. Names and coordinates (Galactic longitude, latitude 
and redshift) of BL Lacs plotted in Fig.~\ref{fig:skymap} which fall
within $3^{\circ}$ from some UHECR event (their energies are listed in 
the last column).}
\begin{tabular}{l|c|c|c|l}
Name & l        &    b    &  z      &  \footnotesize E/$10^{19}$~eV \\ \hline
\footnotesize 1ES 0806+524    &\footnotesize 166.25 &\footnotesize  32.91 %
&\footnotesize  0.138 &\footnotesize 3.4; 2.8; 2.5  \\
\footnotesize RX J10586+5628  &\footnotesize  149.59 &\footnotesize   54.42  %
&\footnotesize   0.144  &\footnotesize  7.76; 5.35  \\ 
\footnotesize 2EG J0432+2910  &\footnotesize  170.52 &\footnotesize  $-12.6$ %
&\footnotesize    -     &\footnotesize  5.47; 4.89  \\
\footnotesize OT 465          &\footnotesize  74.22  &\footnotesize   31.4   %
&\footnotesize    -     &\footnotesize  4.88  \\
\footnotesize TEX 1428+370    &\footnotesize  63.95  &\footnotesize   66.92  %
&\footnotesize   0.564  &\footnotesize  4.97  
\end{tabular}
\end{table}

\begin{figure}
\leavevmode\epsfxsize=3.25in\epsfbox{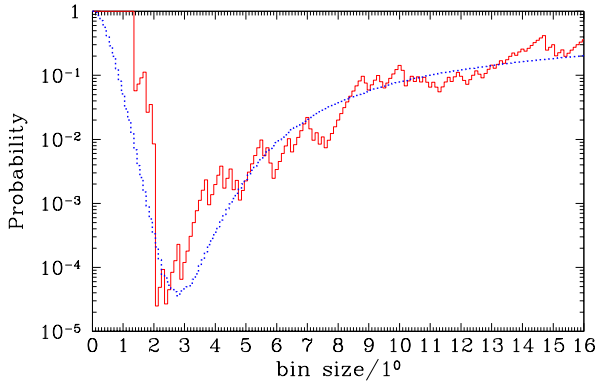}
\caption{FIG. 3. The dependence of the probability $p(\delta)$ on the bin size 
$\delta$ for the combined set of UHECR and BL Lac set (\ref{cuts1}).}
\label{fig:bin2.5}
\end{figure}

The probability $p(\delta)$ as a function of the angular separation
$\delta$ is shown in Fig.~\ref{fig:bin2.5}. It has a minimum at
$2.5^{\circ}$. For comparison, smooth curve shows the behavior
expected when 9 events out of 65 come from BL Lacs (assuming that accuracy
of angle determination is $1.8^{\circ}$ and distribution of
errors is Gaussian ).

The small angular size of the peak in the correlation function,
compatible with the experimental angular resolution, suggests that
UHECR events responsible for these correlations are produced by
neutral primary particles. Indeed, if the primaries were charged they
would have been deflected in the Galactic magnetic field by
$3^{\circ}-7^{\circ}$ depending on arrival direction, particle energy
and the model of the magnetic field, and correlations at $2.5^{\circ}$
would be destroyed.

\paragraph*{Discussion.}

We have seen that 22 bright BL Lacs and 65 cosmic rays from the
combined set are strongly correlated: the probability to find 8 or
more out of 65 randomly generated cosmic rays within $2.5^{\circ}$ of
any of the BL Lacs is $2\times 10^{-5}$. Should one conclude that BL
Lacs are sources of UHECR, or the above correlation may be an artifact
of our selection procedure? Let us discuss possible loopholes.

First potential source of problem is incompleteness of the BL Lac
catalog and non-uniform coverage of the sky. Indeed, 22 BL Lacs
selected by cuts (1) almost all lie in the Northen hemisphere due to
observational bias.  However, it is easy to understand that, unlike
for many other astrophysical problems, for establishing the {\em fact
of correlations} with UHECR the incompleteness of BL Lac catalog is
not essential. The method we use works for any set of potential
sources regardless of their distribution over the sky (including such
extreme cases as just one source, or a compact group of several
sources). This is guaranteed by using the same set of sources with
real data and with each Monte-Carlo configuration.

Second potential problem is related to the fact that there exist
strong autocorrelations in the UHECR set, while Monte-Carlo events are
not correlated. One may wonder if the observed correlation with BL
Lacs is (partially) due to autocorrelations of UHECR. To see that this
effect is negligible in our case, we performed test Monte-Carlo
simulations with configurations containing the same number of doublets
and triplets as the real data, and random in other respects. We found 
practically no difference between the two methods.

Finally, there is an issue of cuts and related issue of selection of
catalogs. One may worry that by adjusting several cuts and searching
in several catalogs the probability as small as $p_{\rm min} \sim
10^{-5}$ can be found with any set of astrophysical objects, even with
those which have nothing to do with UHECR. So, the question is how
easily the low values of $p_{\rm min}$ can be obtained within the
adopted procedure of cuts. This question can be studied quantitatively
by assigning a proper penalty for each try in such a way that
resulting probability gives true measure for the correlations in
question to be a statistical fluctuation. For the case at hand we have
found that when proper penalties are assigned, the resulting
probability is larger than $p_{\rm min}$ by about an order of
magnitude. In other words, one would have to try thousands of catalogs
to find correlation as significant as we have found for BL Lacs.  We
present the procedure of penalty calculation and resulting significance
of correlations below.

In fact, we did not search for correlations with other catalogs of
astrophysical objects. Thus, no penalty is associated with that.
Similarly, we did not adjust the set of cosmic rays (as explained
before, it was selected in Ref. \cite{Tinyakov:2001ic} on the basis of
most significant autocorrelations). But we do adjust cuts in the BL
Lac catalog. Therefore, we should assign a penalty factor to this
adjustment.

It is clear that some cuts have to be made because 65 events may have
at most 65 sources among 306 BL Lacs in the catalog (probably much
less). In our calculations we imposed cuts on redshift, magnitude and
6~cm radio-flux. The cut on redshift is motivated by the expected
total number of sources; we did not adjust this cut to minimize the
probability. Cuts on magnitude and radio-flux were
adjusted. Corresponding penalty can be calculated in the following way
(cf. Ref~\cite{Tinyakov:2001ic}). A random set of cosmic rays should
be generated and treated as real data, i.e. minimum probability
$p_{\rm min}$ is searched for by adjusting the cuts in the BL Lac
catalog in exactly the same way as it was done for the real data.
This should be repeated many times, giving different $p_{\rm min}$
each time. The number of occurrences of a given value of $p_{\rm min}$
is then calculated as a function of $p_{\rm min}$.  This gives the
probability (we call it $p_{\rm cor}$) that the adjustment of the cuts
in BL Lac catalog produces $p \le p_{\rm min}$ with a
random set of cosmic rays. The probability $p_{\rm cor}$ is a correct
measure of the significance of correlations. We define $p_{\rm
cor}/p_{\rm min} > 1$ as the penalty factor.  

We calculated $p_{\rm cor}$ with $10^5$ random sets of cosmic rays. We
have found that the penalty grows at small $p_{\rm min}$ and
approaches a constant value in the limit $p_{\rm min} \rightarrow 0$
(for this reason it is more convenient to define the penalty factor
than to work in terms of $p_{\rm cor}$). For the real set of UHECR
$p_{\rm min} = 4\times 10^{-6}$ and is reached with the cuts
\begin{equation}
\label{cuts2}
z>0.1 \mbox{~or unknown;~~} {\rm mag} <16 ;~~F_6>0.17\mbox{~Jy} 
\end{equation}
They leave 5 BL Lacs two of which coincide with triplets. (In the previous
section different cuts are presented because, with similar significance, they
include more potential sources.) This probability should be multiplied by the
penalty factor. We found that the penalty factor is $\simeq 15$ at
$p_{\rm min} \simeq 10^{-6}$ \cite{binp}. This gives $p_{\rm cor}=6\times 10^{-5}$, which
is the probability that the correlation we have found is a statistical
fluctuation.

\paragraph*{Conclusions. }
The significant correlations between UHECR and BL Lacs imply that at
least some of BL Lacs are sources of UHECR. Most probable candidates
can be seen in Fig.~\ref{fig:skymap} and are listed in Table~1.  Two
BL Lacs, 1ES 0806+524 and RX J10586+5628, coincide with triplets of
UHECR events (in the second case the third event of a triplet is at
$4.5^{\circ}$ and is not listed in the table). Both of them are at the
distance of $\sim 600$~Mpc from the Earth. The next-probable candidate
2EG J0432+2910 has unknown redshift.

The correlations at small angles are difficult to explain by charged
primary particles. Within the Standard Model the only two neutral
candidates are photon and neutrino. Photon attenuation length at $E <
10^{20}$ eV is much smaller (see e.g. \cite{BSigl}) than the distance
to even the closest BL Lac.  However, photons can not be ruled out yet
if one assumes sources at $d \sim 600$ Mpc and ``extreme''
astrophysical conditions: primary particles accelerated to $E >
10^{23}$ eV with ``hard'' spectrum $\sim E^{-\alpha}$ and $\alpha <
2$, and extragalactic magnetic fields $B <10^{-11}$ G \cite{kalashov}.
Neutrino models \cite{neutrinos} require similar assumptions except
that constraints on the magnetic filed are relaxed for ``pure''
neutrino sources and there is no constraint on the distance to the
sources.  However, if ``pure'' neutrino sources cannot be arranged,
the model effectively becomes ``photonic'' \cite{kalashov}.  If
astrophysical difficulties can be overcome, these models will be
appealing candidates for the solution of the UHECR puzzle.
Alternatively, one may resort to a new physics, e.g., violation of the
Lorentz invariance \cite{VLI}.

Independent cross-checks are necessary to determine whether particular
objects are sources of UHECR. One of these cross-checks could be
coincidence of arrival time of events contributing to small angle
correlations with periods of activity of candidate BL Lacs. Dedicated
monitoring of these BL Lac may be suggested. It is also important to
analyze possible specific properties of air showers initiated by these
events.


\section*{Acknowledgments}
{\tolerance=400 We are grateful to S.L.~Dubovsky, K.A.~Postnov,
M.E.~Shaposhnikov and D.V.~Semikoz for valuable comments and
discussions.  This work is supported by the Swiss Science Foundation,
grant 21-58947.99, and by INTAS grant 99-1065. 
}


\begin{thebibliography}{99}

\bibitem{GZK} K. Greisen, Phys. Rev. Lett. \textbf{16}, 748 (1966);\\
  G.T. Zatsepin and V.A. Kuzmin, Pisma Zh. Eksp. Teor. Fiz. \textbf{4}, 144
  (1966) ;

\bibitem{clusters1} X. Chi et al., J. Phys. {\bf G18}, 539 (1992); N. N.
  Efimov and A. A. Mikhailov, Astropart. Phys. {\bf 2}, 329 (1994).

\bibitem{Takeda:1999sg} M.~Takeda {\it et al.}, astro-ph/9902239.

\bibitem{Uchihori:2000gu} Y.~Uchihori {\it et al.}, Astropart.\ Phys.\ {\bf
    13}, 151 (2000) [astro-ph/9908193].

\bibitem{Tinyakov:2001ic} P.~G.~Tinyakov and I.~I.~Tkachev, Pis'ma v ZhETF
  {\bf 74}, 3 (2001), [astro-ph/0102101].

\bibitem{yk} Catalogue of Highest Ehergy Cosmic Rays, No.~3, June 1988, ed.
  A.~Inoue, E.~Sakamoto, World Data Center C2 for Cosmic Rays, Institute of
  Physical and Chemical Research, Wako, Saitama, Japan.

\bibitem{Hayashida:1999zr} N.~Hayashida {\it et al.}, Astrophys.\ J.\ {\bf
    522}, 225 (1999), [astro-ph/0008102].

\bibitem{Dubovsky:2000gv} S.~L.~Dubovsky, P.~G.~Tinyakov and I.~I.~Tkachev,
  Phys.\ Rev.\ Lett.\ {\bf 85}, 1154 (2000), [astro-ph/0001317].

\bibitem{BSigl} P.~Bhattacharjee, G.~Sigl, Phys.Rept.  {\bf 327}, 109 (2000).

\bibitem{correlations} J.~Elbert and P.~Sommers, Astrophys.J.  {\bf 441}, 151
  (1995); G.~Farrar, P.~Biermann, Phys. Rev. Lett.  {\bf 81}, 3579 (1998);
  A.~Virmani, S.~Bhattacharya, P.~Jain, S.~Razzaque, J.P.~Ralston,
  D.W.~McKay., astro-ph/0010235; G.~Sigl, D.~F.~Torres, L.~A.~Anchordoqui and
  G.~E.~Romero, Phys.\ Rev.\ D {\bf 63}, 081302 (2001), [astro-ph/0008363];
  S. S. Al-Dargazelli et al, J. Phys.  G {\bf 22}, 1825 (1996); G. R. Farrar
  and T. Piran, astro-ph/0010370; A. A. Mikhailov, in Proceedings of the 26th
  International Cosmic Ray Conference, Salt Lake City, 1999, Vol. 3, p. 268;
  A. A. Mikhailov and E. S. Nikiforova, JETP Letters, {\bf 72}, 229 (2000).

\bibitem{catalogue} M.P. Veron-Cetty and P. Veron, Quasars and Active Galactic
  Nuclei, ESO Scientific Report 9th Ed. (2000).

\bibitem{binp} This number does not include penalty for the adjustment of the
  bin size because minimum of $p(\delta )$ is consistent with the angular
  resolution.  Corresponding factor is $\approx 4$.
  
\bibitem{kalashov} O.~E.~Kalashev, V.~A.~Kuzmin, D.~V.~Semikoz and
  I.~I.~Tkachev, [astro-ph/0107130].
  
\bibitem{neutrinos} T. J. Weiler, Phys. Rev. Lett. {\bf 49}, 234 (1982); D.
  Fargion, B. Mele, A. Salis, ApJ {\bf 517}, 725 (1999); S.~Yoshida, G.~Sigl
  and S.~Lee, Phys.\ Rev.\ Lett.\ {\bf 81}, 5505 (1998); J.~J.~Blanco-Pillado,
  R.~A.~Vazquez and E.~Zas, Phys.\ Rev.\ D {\bf 61}, 123003 (2000); G.~Gelmini
  and A.~Kusenko, Phys.\ Rev.\ Lett.\ {\bf 82}, 5202 (1999), [hep-ph/9902354].

\bibitem{VLI} S.~R.~Coleman and S.~L.~Glashow, Phys.\ Rev.\ D {\bf 59}, 116008
  (1999) [hep-ph/9812418];
  S.~L.~Dubovsky and P.~G.~Tinyakov, [astro-ph/0106472].

\end{thebibliography}
\end{document}